\documentstyle[12pt]{article}
\begin{document}
\begin{titlepage}
%\hfill{SNUTP 92-57}
\vskip 1.5cm
\begin{center}
{\Large\bf Exclusive heavy meson pair production by\\
 $\gamma\gamma$ collision in  heavy quark effective theory }
\end{center}
\vskip 2cm
\begin{center}
{S. Y. Choi\footnote{E-mail : SYCHOI@KRSNUCC1.bitnet} and H. S. Song}
\end{center}
\vskip 0.5cm
\begin{center}
{\it Center for Theoretical Physics and Department of Physics,\\
Seoul National University, Seoul 151-742, Korea}
\end{center}
\vskip 5cm
\begin{center}
{\bf ABSTRACT}
\end{center}
The spin and flavor symmetries of the heavy quark effective theory are
employed to discuss the exclusive cross sections for pair production
of heavy mesons in photon-photon collision. The ratios of the exclusive
cross sections for heavy mesons are obtained explicitly and
the validity of results is discussed.
\end{titlepage}
\newpage
\section{Introduction}
\label{sec:intro}

Physical processes of heavy mesons involving heavy quarks,
such as the $c$ and $b$ quarks, have recently attracted
both theoretical and experimental interests.

The experimental data have improved
considerably and also proposals \cite{Feld} for detailed study of such
processes have been made. On the other hand, new theoretical ideas
developed recently have led to the formulation of an effective
heavy quark theory \cite{Volo,Isgu,Grin,Falk}
with which the analysis of physical processes is
greatly simplified due to additional symmetries arising in the
heavy quark limit where the heavy quark mass is very large.

One of the symmetries is a heavy flavor symmetry, namely an
SU($N$) for $N$ heavy flavors, under which the heavy quarks
are rotated into one another.
The second is called the spin symmetry that arises due to the
decoupling of the spin degrees of freedom in the heavy quark
limit.
And one \cite{Georgi} of the important properties of the heavy qaurk
effective theory is that the heavy quark velocity is difficult
to change as long as
nonperturbative aspects of QCD with a typical
scale $\Lambda_{\rm QCD}$ are concerned
but it can be only modified by perturbative processes
such as an electroweak interaction.

These symmetries and the so-called velocity superselection rule
have been very successful in their applications to weak decays of heavy
hadrons \cite{Isgu,Mann} and exclusive heavy hadron production
in $e^+e^-$ annihilation \cite{Falk1,Mann1,Mann2}.
While the additional symmetries are broken by terms of the
order $\Lambda_{\rm QCD}/m_H$, where $m_H$ is the mass of the
heavy meson involved in the particular process and by QCD
radiative corrections of the order $\alpha_S(m_H)$, these
corrections \cite{Falk,Falk1}
may be treated systematically and have been calculated
in some cases.

In this paper, we use symmetries of the heavy quark effective
theory to discuss pair production of heavy mesons in
$\gamma\gamma$ collision.

The cross section for production of a pair of heavy mesons in
two-photon reactions is a function of the dimensionful variables
$s=(k_1+k_2)^2$, $m^2_H$, $\Lambda^2_{\rm QCD}$ and $Q^2$,
which is a generic notation for
the absolute value of momentum transfer due to
$t$- and $u$-channel heavy quark exchanges not involved
in $e^+e^-$ collisions.
We consider the kinematic regime where $\sqrt{s}$ is reasonably
larger than $m_H$, and both $m^2_H$ and $Q^2$ are much larger
than $\Lambda^2_{\rm QCD}$. Then the pair production of heavy mesons
in the $\gamma \gamma$ (as well as $e^+e^-$) collisions may be
visualized as a heavy quark pair production followed by
the fragmentation of the heavy quark into a heavy meson involving
the light QCD degrees of freedom.
The approximate visualization may get more concrete with
an appropriate  angular cutoff to the production distribution
of heavy mesons.

The scattering amplitudes of large-angle exclusive processes
\cite{Brod} might be computed, in the leading order, from Born
diagrams shown in figs.\ \ref{Feynman}(a-h).
The solid line is for the heavy quark and the dashed line for
the light degree of freedom in the heavy meson.
However, for the heavy meson containing a heavy quark,
the scattering amplitudes of figs.\ \ref{Feynman}(c-h) are strongly
suppressed, since the diagrams require the exchange of a hard gluon to
produce the final state, while those of figs.\ \ref{Feynman}(a) and
\ref{Feynman}(b) do not.
Therefore, in the heavy quark limit, only the two diagrams (a) and (b)
contribute to the production of a pair of heavy mesons.

Basically, the two photon system \cite{Olsson}
is distinctly different from the $e^+e^-$ system
in the following aspects.
First, the former has positive charge conjugation, while
the latter has negative one. Thus the C conservation of both QED
and QCD guarantees a separate and complementary investigation
of final meson states by $\gamma\gamma$ and $e^+e^-$ collisions.
On the other hand, virtual photons are also available at $e^+e^-$
or $pp(\bar{p})$ colliders so that spin-one final states can
be explored.
Finally, the on and off shell nature and the polarization of
each of the (virtual or real) incident photons can be
continuously varied, allowing detailed tests of the theory.
The last two aspects are not considered and only on-shell photons
are taken into account in the present work.

The paper is organized as follows. In section~2, we determine
heavy meson wave functions through symmetries of the heavy
quark effective theory and apply them to obtain the exclusive
cross sections for the production of heavy mesons in $\gamma\gamma$
collision and their ratios. In section~3, our results are discussed
and some conclusions are made.

\section{Cross sections and ratios}
\label{sec:cross section}

We employ the heavy quark effective theory to determine
the wave functions of heavy mesons formed by a non-tractable
fragmentation mechanism from heavy quarks and to apply them to
the production mechanism of heavy mesons, especially, in
the $\gamma\gamma$ collision.

In general, a heavy meson containing a single heavy quark
is mostly viewed as a freely propagating
point-like color source (a heavy quark), dressed by strongly
interacting light degrees of freedom (so-called ``brown mucks")
bearing appropriate values of color, baryon number, angular
momentum and parity to make up the observed physical state.
The heavy meson is described by the velocity
vector $v^\mu$ ($v^2=1,v^0>0$) and the residual momentum vector
$k^\mu$, which form the total four momentum
$p^\mu (=m_H v^\mu+ k^\mu)$.
The velocity $v^\mu$ is prohibited from changing by the velocity
superselection rule and $k^\mu$ is of the order $\Lambda_{\rm QCD}$.
Moreover heavy quark states and heavy  anti-quark states are
completely decoupled because of an unsurmountable mass gap
$2m_H$ and thus, in each velocity sector of the theory, there are
two separate SU(2) spin symmetries to enable us to connect
different spin states of heavy mesons and those of heavy
anti-mesons, respectively.

These spin symmetries can be used to determine wave functions of
heavy mesons and their antimesons explicitly \cite{Geor1,Falk3}
and to derive relations between the matrix elements of
any Dirac structure $\Gamma$ which is to be sandwiched between
heavy quark fields.
The Dirac structure $\Gamma$ denoting the production of
a pair of heavy quarks through $t$- and $u$-channel heavy quark
exchanges in $\gamma\gamma$ collision is given by
\begin{eqnarray}
\Gamma=\not\!{\epsilon_1}
       \frac{m_H (1+\not\!{v_1})-\not\!{k_1}}{v_1\cdot k_1}
       \not\!{\epsilon_2}
      +\not\!{\epsilon_2}
       \frac{m_H (1+\not\!{v_1})-\not\!{k_2}}{v_1\cdot k_2}
       \not\!{\epsilon_1},\label{operator}
\end{eqnarray}
where $\epsilon_1$ and $\epsilon_2$ are wave vectors of the incident
photons with  four momentum vectors $k_1$ and $k_2$,
respectively, and $v_1(v_2)$ is the four velocity of the
heavy meson (antimeson). The first term in eq.\ (\ref{operator})
corresponds to the $t$-channel exchange and the second to the $u$-channel
exchange.

A multiplet of heavy mesons, transformed into one another by the
spin symmetry, should have the same mass $m_H$ and thus their genuine
mass differences at most of the order $\Lambda_{\rm QCD}$ are
neglected in the heavy quark limit.

The matrix wave functions of heavy mesons containing light degrees of
freedom of zero orbital angular momentum as well as complying with
Lorentz invariance and parity are as follows:
\begin{eqnarray}
M_v &=& \frac{1+\not\!{v}}{2}\gamma_5 \ \ : \ \
        {\rm a\ \ pseudoscalar\ \ meson},\nonumber\\
M^*_v &=& \frac{1+\not\!{v}}{2}\not\!{\epsilon} \ \ : \ \
          {\rm a\ \ vector\ \ meson},\nonumber\\
\tilde{M}_v &=& \frac{1-\not\!{v}}{2}\gamma_5 \ \ : \ \
                {\rm a\ \ pseudoscalar\ \ antimeson},\nonumber\\
\tilde{M^*_v} &=& \frac{1-\not\!{v}}{2}\not\!{\epsilon} \ \ : \ \
                  {\rm a\ \ vector\ \ antimeson}.
\end{eqnarray}
The complex conjugates $\bar{M}_v$ and $\bar{M}^*_v$ are
defined as
\begin{eqnarray}
\bar{M}_v = \gamma^0 M^+_v \gamma^0,\ \
\bar{M}^*_v = \gamma^0 {M^*_v}^+ \gamma^0,
\end{eqnarray}
respectively.
Then the matrix element of two heavy mesons may be expressed by just
one form-factor \cite{Falk1,Mann1} as
\begin{eqnarray}
<M_{v_1}\tilde{M}_{v_2}|\bar{h}\Gamma\tilde{h}|0>=
m_H \zeta(v_1\cdot v_2){\rm tr}(\bar{M}_{v_1}\Gamma\tilde{M_{v_2}}),
\end{eqnarray}
where $h$ ($\tilde{h}$) is the heavy quark (anti-quark) field,
respectively, and $\zeta(v_1\cdot v_2)$ is the analogue of the
dimensionless Isgur-Wise function for the time-like momentum transfer
$v_1\cdot v_2$, which is in the center of mass frame
\begin{eqnarray}
v_1\cdot v_2=\frac{s}{2m^2_H}-1.
\end{eqnarray}

Note that only one Isgur-Wise form factor is needed for heavy
mesons.
Therefore, the ratios of the exclusive production of the
$M\tilde{M}$, $M^*\tilde{M}(M\tilde{M}^*)$ and $M^*\tilde{M}^*$
meson pairs in the $\gamma\gamma$ collision will be independent
of the unknown coefficient function $\zeta(v_1\cdot v_2)$.

For simplicity, let us introduce a variable $z$ as
\begin{eqnarray}
z=\beta\cos\theta,
\end{eqnarray}
where $\theta$ is the scattering angle and
$\beta=\sqrt{1-4m^2_H/s}$ in the center of mass frame.
The general form of differential cross section for
$\gamma\gamma\rightarrow M\tilde{M}$,
where $M$ is a heavy meson, is given by
\begin{eqnarray}
\frac{{\rm d}\sigma}{{\rm d}z}(\gamma\gamma\rightarrow
 M\tilde{M})=\frac{\pi \alpha^2}{8}N_Cq^4_H|\zeta(v_1\cdot v_2)|^2
 \frac{1}{s}|{\rm tr}(\bar{M}_{v_1}\Gamma\tilde{M}_{v_2})|^2,
\end{eqnarray}
where  $q_H$ is the charge
of a heavy quark and  $N_C$ is the color factor which is 3 for
${\rm SU(3)}_{\rm C}$.
Taking an average over initial photon polarizations and a sum over
final vector meson polarizations, we find for the differential cross
sections
\begin{eqnarray}
&&\frac{{\rm d}\sigma}{{\rm d}z}(\gamma\gamma\rightarrow M\tilde{M})
  =\sigma_0\times\frac{f(\beta,z)}{(1-z^2)^2},\nonumber\\
&&\frac{{\rm d}\sigma}{{\rm d}z}
  (\gamma\gamma\rightarrow M^*\tilde{M}(M\tilde{M}^*))
  =\sigma_0\times\frac{s}{8m^2_H}
   \frac{2z^2-3z^4+f(\beta,z)}{(1-z^2)^2},\nonumber\\
&&\frac{{\rm d}\sigma}{{\rm d}z}
  (\gamma\gamma\rightarrow M^*\tilde{M}^*)
  =\sigma_0\times
   \frac{\frac{s}{2m^2_H}(\beta^2-z^4)+3f(\beta,z)}{(1-z^2)^2},
   \label{differential}
\end{eqnarray}
where
\begin{eqnarray}
&&\sigma_0=\pi\alpha^2N_Cq^4_H
           \frac{|\zeta(v_1\cdot v_2)|^2}{s},\nonumber\\
&&f(\beta,z)=\beta^4+(z^2-\beta^2)^2.
\end{eqnarray}
The ratios of differential cross sections are
$\beta^2:\frac{s}{8m^2_H}\beta^2:\frac{s}{4m^2_H}+\frac{3}{2}\beta^2$
at $\theta=\frac{\pi}{2}$ and $\beta^2:1:2+3\beta^2$ at the forward
or backward direction ($\theta=0$ or $\pi$), respectively.
The production of two vector mesons dominates in both cases.

Integrating the differential cross sections over $z$ we get the ratios
\begin{eqnarray}
&&\sigma(\gamma\gamma\rightarrow M\tilde{M}):
  \sigma(\gamma\gamma\rightarrow M^*\tilde{M}):
  \sigma(\gamma\gamma\rightarrow M^*\tilde{M}^*)\nonumber\\
&&\hskip 3cm =h_1(\beta):\frac{s}{8m^2_H}h_2(\beta):
   3h_1(\beta)+\frac{s}{4m^2_4}h_3(\beta),\label{total}
\end{eqnarray}
where
\begin{eqnarray}
h_1(\beta)&=&\frac{\beta}{1-\beta^2}(2\beta^4-4\beta^2+3)
           +(\beta^4+\beta^2-\frac{3}{2})
           {\rm log}\frac{1+\beta}{1-\beta},\nonumber\\
h_2(\beta)&=&\frac{(1+\beta^2)^2}{2}[{\rm log}\frac{1+\beta}{1-\beta}
             -\frac{2\beta}{1+\beta^2}],\nonumber\\
h_3(\beta)&=&-6\beta+(\beta^2+3){\rm log}\frac{1+\beta}{1-\beta}.
\end{eqnarray}
Near the threshold $\sqrt{s}=2m_H$, the $\beta$ value is small
enough to be used as an expansion parameter.
As $\beta\rightarrow 0$, the pseudoscalar meson pair production is strongly
suppressed and the ratios of other total cross sections come close to
$1:6$ (See fig.\ \ref{ratios}).

\section{Discussion}
\label{sec:discussion}

In this paper we have obtained the ratios (\ref{total}) of the exclusive
cross section in the limit where effects of the order of
$\Lambda_{\rm QCD}/m_H$ and $\alpha_S(m_H)$ may be
neglected and momentum transfers are large compared to
$\Lambda^2_{\rm QCD}$. We find that near the threshold
$\sqrt{s}=2m_H$ the process $\gamma\gamma\rightarrow M\tilde{M}$
is highly suppressed and the ratios of other total cross sections become $1:6$.
In other words, the vector meson production is dominant and
any pseudoscalar meson should be accompanied by a vector meson
as its production companion in the $\gamma\gamma$ collision.
The result is quite different from that \cite{Falk1,Mann1} in the $e^+e^-$
case where the ratios are $1:2:7$ at threshold.

{}From eq.\ (\ref{differential}),
we can show that  the differential cross sections
near threshold are
\begin{eqnarray}
\frac{{\rm d}\sigma}{{\rm d}z}(\gamma\gamma\rightarrow M^*\tilde{M})
\propto \cos^2\theta,\ \ \hskip 0.5cm
\frac{{\rm d}\sigma}{{\rm d}z}(\gamma\gamma\rightarrow M^*\tilde{M}^*)
\propto 2.
\end{eqnarray}
The angular dependence implies :
\begin{itemize}
\item{The final peudoscalar and vector meson system has
       a {\it P}-state wave function $Y_{10}(\cos\theta)$.}
\item{The final vector meson pair system has an isotropic
      distribution, i.e. no angular momentum.}
\end{itemize}
Naturally the ratios are expected to be $1:6$ near threshold, which
is consistent with the explicit calculation.

Our result has a finite region of validity. It is quite likely that,
at the low end around the threshold, important resonance effects
arise and influence the ratios.
Thus our prediction would work only in the energy region at least
a few GeV higher than $\sqrt{s}=2m_H$.

The upper limit of validity is  complicated.
We should consider  QCD radiative corrections
as well as effects of higher order in $\Lambda_{\rm QCD}/m_H$.
The latter corrections \cite{Falk1,Falk3}
can be systematically treated. However, the former are difficult
to treat, because of non-perturbative QCD contributions.
Besides, the angular distributions are forwardly and backwardly peaked
and the cross sections become too small at high energies to verify
any theoretical prediction.

Consequently, the appropriate choice of the region of validity
is highly required. In the $\gamma\gamma$ production of heavy mesons
the appropriate region of validity is the one where
the CM energy $\sqrt{s}$ is a couple of GeV higher than
$2m_H$ but not so very high for data analyses, and
the $t$- or $u$-channel momentum transfer $Q^2$
should be much larger than $\Lambda^2_{\rm QCD}$
with a view to excluding non-perturbative QCD effects in the heavy
quark pair production by two photon before hadronization.

\begin{flushleft}
{\Large\bf Acknowledgement}
\end{flushleft}

This work was supported in part by the Korea Science and Engineering
Foundation and in part by the Korean Ministry of Education.

\newpage
\begin{flushleft}
{\Large\bf Figure captions}
\end{flushleft}
\begin{figure*}[h]
\caption{Feynman diagrams contributing to heavy meson pair production
        in $\gamma\gamma$
        collision to lowest order in $\alpha_S$
        at the QCD parton level. The solid line is for the
        heavy quark and the dashed line for the light quark.}
        \label{Feynman}
\caption{Ratios of total cross sections
        $\sigma (\gamma\gamma\rightarrow M\tilde{M})/
         \sigma (\gamma\gamma\rightarrow M^*\tilde{M})$ and
        $\sigma (\gamma\gamma\rightarrow M^*\tilde{M}^*)/
         \sigma (\gamma\gamma\rightarrow M^*\tilde{M})$
        versus the heavy meson velocity $\beta$.}
        \label{ratios}
\end{figure*}
\end{document}